\documentclass[conference]{IEEEtran}
\IEEEoverridecommandlockouts
\usepackage{cite}
\usepackage{amsmath,amssymb,amsfonts}
\usepackage{algorithmic}
\usepackage{graphicx}
\usepackage{textcomp}
\usepackage{xcolor}
\def\BibTeX{{\rm B\kern-.05em{\sc i\kern-.025em b}\kern-.08em
    T\kern-.1667em\lower.7ex\hbox{E}\kern-.125emX}}
\begin{document}

\title{Negative-ResNet: Noisy Ambulatory Electrocardiogram Signal Classification Scheme}

%\thanks{The authors would like to acknowledge the support of an ARC DP grant.}

% \author{\IEEEauthorblockN{Zihuai Lin}
% \IEEEauthorblockA{\textit{$^*$School of Electrical and Information Engineering, The University of Sydney, Australia} \\
% {zihuai.lin{@sydney.edu.au}}}
% }

\author{\IEEEauthorblockN{Zijiao Chen$^*$, Zihuai Lin$^*$, Peng Wang$^*$, Ming Ding$^+$}
\IEEEauthorblockA{\textit{$^*$School of Electrical and Information Engineering, The University of Sydney, Australia} \\
\textit{$^+$Data61, CSIRO, Australia} \\
{zche8039@uni.sydney.edu.au};{\rm{\{}zihuai.lin, thomaspeng.wang\rm{\}}{@sydney.edu.au}}; {ming.ding@data61.csiro.au}}
}

\maketitle

\begin{abstract}
With recently successful applications of deep learning in computer vision and general signal processing, 
deep learning has shown many unique advantages in medical signal processing. However, 
data labelling quality has become one of the most significant issues for AI applications, 
especially when it requires domain knowledge (e.g. medical image labelling).
In addition, 
there might be noisy labels in practical datasets, 
which might impair the training process of neural networks.
In this work, 
we propose a semi-supervised algorithm for training data samples with noisy labels by performing selected Positive Learning(PL) and Negative Learning(NL). 
To verify the effectiveness of the proposed scheme, 
we designed a portable ECG patch - iRealCare and applied the algorithm on a real-life dataset. 
Our experimental results show that we can achieve an accuracy of 91.0\,\%, which is 6.2\,\% higher than a normal training process with ResNet. 
There are 65 patients in our dataset and we randomly picked 2 patients to perform validation. 

\end{abstract}
 
This is a post-peer-review, pre-copyedit version of an article published in Neural Computing and Applications. The final authenticated
version is available online at: https://doi.org/10.1007/s00521-020-05635-7”.

\section{Introduction}

Machine learning and deep learning have become the most popular data analysis techniques over these years due to its high accuracy, reliability and repeatable results. 
Especially in the field of biomedical image processing~\cite{r8}, 
where these cutting-edge techniques can be used to  assist medical professionals with their diagnosis. 
In traditional medical imaging, 
diagnosis mainly relays on human eyes and experiences, 
which can sometime be inconsistent and problematic. 
Neural network is proved to possess an extraordinary ability to recognize patterns and make classifications.
It is able to capture high-level features of signals and images, 
which are very subtle or even impossible for human eyes. 
As such, 
neural network is a powerful tool for solving biomedical image processing problems and is widely used over the past few years.

However, 
neural network, 
a data-driven method, 
relies heavily on the training dataset, 
which will lead to unexpected performance if it does not have a decent size or proper label~\cite{r25}. 
In many medical research works, 
it is challenging to obtain a dataset with a huge size and labelling them correctly due to many factors(e.g. cost, domain knowledge). 
 
Labelling such a large dataset tends to be erroneous due to human mistakes and medical misjudgment. 
It will therefore need more quality control (QC) procedures such as reliability cross-check~\cite{r26}. 
And these will not always be possible as QC can double the number of data that needs to be labelled. 

In this work, 
to overcome these issues and design a more robust training scheme that can be used for data classification 
when the biomedical signal is collected under an imperfect condition, 
we proposed a \textbf{Negative-ResNet} with 5 residual blocks (12-layer Convolutional Neural Network in total), 
which can be trained and showed good performance even when the dataset contains a significant number of \textbf{noisy labels}.

\begin{figure}[h]
\centering
\includegraphics[width=8cm]{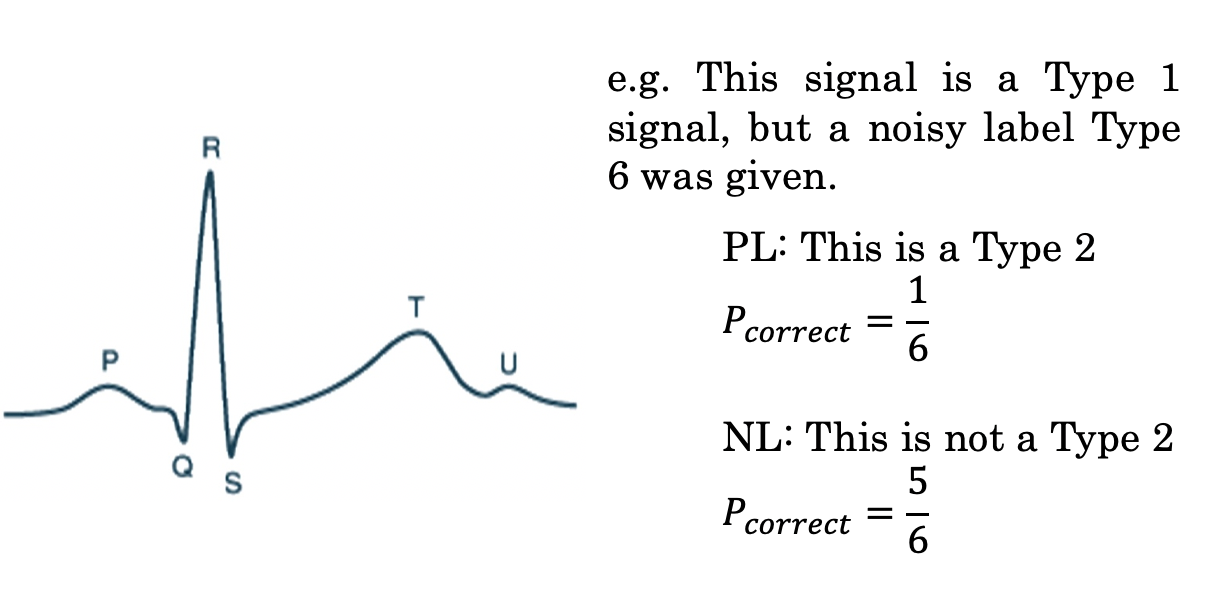}
\caption{Negative Learning(NL) versus. Positive Learning(PL), 
NL is more likely to provide the correct information}
\label{fig_NLPL}  
\end{figure}

Unlike usual Positive Learning (PL), 
the \textbf{Negative Learning (NL)} training scheme we mention above provides the information that a given signal doesn't belong to a  certain label, 
which is more likely to be correct.
For example, 
in Fig. \ref{fig_NLPL}, 
we plot 6 types of data in total in the dataset. 
If we assume a label might be noisy:
For positive learning, 
information "This is a Type 2" will be provided,
As such, there's only 1/6 chance that the label is correct. 
However, for negative learning, information "This is not a Type 2" will be given. 
The probability of correctness increases to 5/6.
Hence, NL will usually provide more precise information when the label is noisy.

To verify the effeteness of the training scheme, 
we designed a portable Electrocardiogram (ECG) patch
- iRealCare 
to collect and construct our ambulatory ECG dataset.  
There are 5 types of signals in this dataset, 
which indicates 5 different status of the heartbeat signal. 
And the size of the dataset is more than 4 million of heartbeat signals.

The main contributions of this work are listed as below:

\begin{itemize}
\item We \textbf{propose a novel semi-supervised training scheme}: 
a confidence-level-based training method, 
following by a negative training scheme using a deep residual block based neural network backbone.
\item We develop a \textbf{novel portable IoT ECG Patch} named iRealCare and constructed a \textbf{large real-patient dataset} using this IoT device with more than 4 million ECG signal waveforms.
\item We apply the proposed scheme on our dataset and achieve a satisfactory\textbf{ classification accuracy of 91.0\,\%}, 
which is \textbf{6.2\,\% higher than a normal training process} with ResNet.
\end{itemize}

The remainder of this manuscript is organized as below: 
Section \ref{related work} mainly talks about the previous relevant study related to our work, 
such as ECG signal and noisy label training.
Section \ref{s2} describes how our dataset is constructed and issues that comes with it. 
It mainly introduces the ECG patch, the dataset and the data pre-processing procedure. 
Section \ref{s3} demonstrates the methodology of the project, 
including detailed explanations of the concept of negative learning and confidence-level-based learning; the novel structure of the residual block.
Section \ref{s4} introduces our proposed deep learning architecture and the experiments for performing and evaluating the proposed method.
Section \ref{s5} gives analytical results of the procedure.
In the end, Section \ref{s6} concludes the work.

\section{Related Work\label{related work}}
\subsection{ECG Signal}
Arrhythmia, 
also known as irregular heartbeat, 
is an important indicator for most of the heart diseases~\cite{r1}. 
Many infrequent arrhythmias are harmless while some may cause serious damage or even death~\cite{r1}. 

\begin{figure}[h]
\centering
\includegraphics[width=6cm]{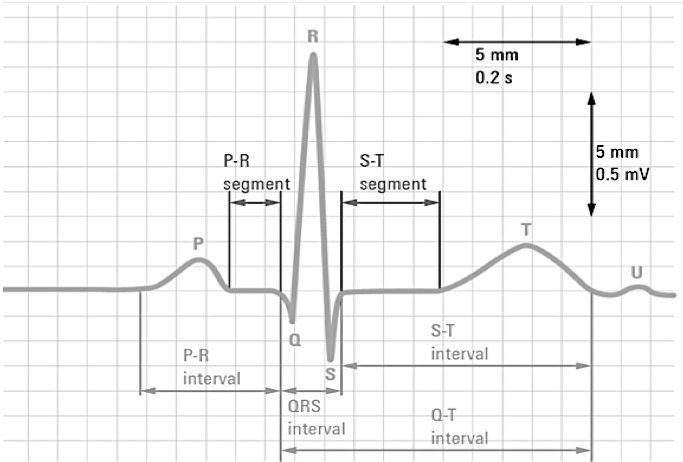}
\caption{Standard waveform of an Electrocardiogram}
\label{fig_1}
\end{figure}

Electrocardiogram(ECG) is currently widely used for detecting arrhythmia by measuring the electrical activities generated by the heart as contracts~\cite{r1,r2,r3,r4}. An ECG signal can be characterized by two features: 
peaks and intervals. 
Each peak (P, Q, R, S, T) has its normal amplitude, 
and each interval (P-R, R-R, Q-R-S, S-T) has its duration period. 
Fig. \ref{fig_1} illustrates an ECG signal, 
and Table \ref{TB:tb_1} shows the ECG features and their typical values\cite{r5,r6}.

\begin{table}[h]
\caption{Feature value of normal ECG}
	\begin{center}
		\begin{tabular}{|c|c|}
			\hline
			Feature&Value\\
			\hline
		    PR interval& 0.12-0.20s\\
			P-wave interval&0.00-0.12s\\
			QRS&0.06-0.11S\\
			RR interval&0.60-1.00s\\
			R-wave amplitude& $>$ 0.50mV\\
			QT interval& 0.33-0.43s\\
			T-wave amplitude& $>$ 0.50mV\\
			P-wave amplitude& $<$ 0.25mV\\
			\hline
		\end{tabular}
	\end{center}
	\label{TB:tb_1}
\end{table}

However, ECG signal data are usually of large size, 
and it would be a huge workload for cardiologists to 
review all of them manually. 
Hence, 
over the past few decades, 
many researchers and engineers have devoted themselves to design robust and accurate equipment and algorithms for ECG based arrhythmia detection and classification. 

Conventionally, 
ECG decomposition and hand-crafted manual features are used for ECG classification. 
In~\cite{r7},~\cite{r8} and~\cite{r9}, 
the wavelet transform was used to extract features of each QRS complex. In~\cite{r10} and~\cite{r11}, adaptive filtering was designed and used for arrhythmia detection. 

Even though good results were reported, 
these model-based feature extraction methods are not generic and robust. 
Individual differences in ECG signals are significant and may lead to poor performance for different individuals. 

Neural networks which are able to extract high-level features are proved to be a powerful tool to learn distributions of a dataset and generalize from training data~\cite{r12,r13}. 
And the authors of~\cite{r14} and~\cite{r15} both presented a machine learning based method which consisted of a 1-D Convolutional Neural Network (CNN) for real-time ECG classification.~\cite{r15} used a nine-layer CNN to classify five classes of arrhythmia and reached an accuracy of 94\% on MIT-BIH dataset. Even though~\cite{r15} reached a very high accuracy, it failed to consider performance on individual patients.~\cite{r14} presented a better way of introducing patient-specific training using transfer learning. And it is also based on MIT-BIH database.  

Besides CNN, other neural network backbone models are also commonly used in ECG classification. In~\cite{r16}, a 3-layer Recurrent Neural Network was used for 5-type classification and an 85\% of accuracy was performed. In~\cite{r17,r18,r19}, long short-term memory(LSTM) showed very promising result in arrhythmia detection and classification tasks. 

Even though~\cite{r10},~\cite{r14},~\cite{r7},~\cite{r8} and~\cite{r12} showed very significant improvement in classification than using naked eyes, 
they suffer from either patient-generation ability or adaptability to the real-world data. 

In~\cite{r20} and~\cite{r21} , 
patient-specific ECG classifiers were built using transfer learning of 2-D convolutional neural networks. 
A convolutional neural network was trained and used for feature extraction first. 
Then an individual classifier was trained for each patient transferring the whole pre-trained model from the training set domain to a specific patient domain. 

However, 
these researches were all built on the MIT-BIH dataset, 
which is a very classic public and wholesome dataset~\cite{r22}. 
The data is clean and collected from a 24-hour monitoring Holter~\cite{r23,r24}, 
which has a significant difference from the data collected by an ambulatory IoT patch. 
In many Smart Health applications involving ECG, 
the collected data will not be as clean and high-quality as the MIT-BIH dataset and therefore requires extra data analysis.

\subsection{Noisy Label Training}

In the era of big data, obtaining large-scale dataset with high quality and precise labels is one of the most important procedures for most of data analysis related projects, 
especially when data-driven methods such as machine learning and deep learning are applied. 
However, annotating such a large dataset usually requires tremendous human resources and sometimes even requires domain knowledge. 
As such, human mistakes occur very often due to the reasons mentioned above, which will lead to a dataset with a certain number of noisy labels. 
To solve this problem, numerous studies are done in the field of noisy label training in the past several decades. 
And here, we refer to this survey~\cite{01} as a comprehensive review.

Many recent studies have illustrate directly training the noisy labels will impair the training scheme and therefore negatively impact the classification accuracy~\cite{02}. To avoid this issue, there are 2 major types of noisy label training schemes: Noisy Model Based Methods and Noisy Model Free Methods.

\subsubsection{Noisy Model Based Methods}

Some approaches advocate to use noise model based methods which mainly aims to model the noise structure and extract the noise feature, 
so that the information of the noise can be estimated and applied on the training scheme. The performance of this method usually depends on the accuracy of noisy estimation.
Hence, it's more suitable when prior knowledge about noise structure is known.

In~\cite{03, 04, 05, 07, 08}, noise channels were modelled at the training phase and the noise got removed at the evaluation phase. Even through good results were performed, the most significant problem in this method is the scale-ability. The calculation would become intractable when the number of class increases in classification task.

Some approaches attempted to identify and clean noisy labels by picking the suspicious samples and send them to human for further annotation~\cite{09, 10}. However, this could still be not feasible when the dataset is at a large size.

Other approaches including dataset pruning~\cite{11, 12,13}, sample choosing~\cite{14, 15,16,17}, sample importance weighting~\cite{18, 19,20,21} also show significant improvement on noisy data classification. However, all of these methods are highly dependent on the calculation and estimation of the noisy structure. As such, prior information tends to be essential in this case.

\subsubsection{Noisy Model Free Methods}

On the other hand, instead of building a noise structure model, 
noisy model free methods tends to achieve label noise robustness.
These methods requires that the robustness of the classifier, 
which cannot be too sensitive  to noise.

Some methods aimed to design a robust loss function so that the noisy would not decrease the accuracy~\cite{22,23,24,25}. Many researchers proposed different type of loss function which showed good result, such as distance-based loss~\cite{26}, information-theoretic loss~\cite{27},  classification-calibrated loss functions~\cite{28}.

With the development of deep neural networks, researchers tend to transfer from hand-designed features to autonomously learned machine learning techniques. So does in noisy label training, a machine learning algorithm called Meta Learning is widely used to learn the high level features on their own~\cite{29,30,31}.

In addition, there are also some other techniques such as regularizer~\cite{32,33,34}, which was used to prevent overfitting from  noisy labels. Besides, complementary labels~\cite{35} and prototype learning~\cite{36} were used for noisy label training as well.

\section{ECG Data Acquisition, Processing and Accuracy considerations}\label{s2}

\subsection{iRealCare ECG patch}
\begin{figure}[h]
\centering
\includegraphics[width=6cm]{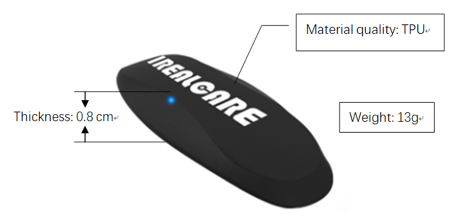}
\caption{iRealCare IoT patch}
\label{fig_2}
\end{figure}

Our dataset was collected by a device named iRealCare~\cite{r27} as shown in Fig.\ref{fig_2}. 
It’s a portable ECG monitoring device with signals sampled at 250Hz. 
The patch is lightweight with only 13 grams. 
In addition, 
the patch has the characteristic of being able to perform long-term continuous monitoring, 
which can last 72 hours. 
As such, 
it has significant advantages over the traditional ECG Holter.

\subsection{Dataset}

Our experiments were performed on an ECG signal dataset collected from 65 patients.
4,831,517 heartbeats were identified considering the position of QRS, 
among which heartbeats were divided into 6 categories, 

Normal Heartbeat(N); Ventricular premature beat (V); Supraventricular premature beat(S); Atrial fibrillation(A); Electromagnetic Interference(E) and Motion Interference(Q).
In order to construct a balanced dataset, 
each category contains the same number of data points. 
Therefore, 
no bias will be introduced to the neural network during training because of the different numbers of training samples for each category.
In other words, 
each category is equally represented by data samples, 
which avoids the over-fitting problem for certain categories. 
Table~\ref{tab_2} shows the number of six types of ECG signals.

\begin{table}[h]
\caption{Number for six types of ECG signals}
	\begin{center}
		\begin{tabular}{|c|c|c|}
			\hline
			Type&Meaning&number\\
			\hline
			N&Normal Heartbeat& 3,501,003\\
			V&Ventricular premature beat&	194,469\\
			S&	Supraventricular premature beat&45,536\\
			A&	Atrial fibrillation&96,052\\
			E& Electromagnetic Interference& 46,437\\
			Q&	Movement Interference& 148,020\\
			\hline
		\end{tabular}
	\end{center}
	\label{tab_2}
\end{table}

As mentioned before, 
the dataset generated by the patch would be more noisier than the Holter due to the limitation of the IoT patch, which could generate a significant number of Electromagnetic Interference. 
In addition, 
a considerable amount of movement interference will be introduced when the patients wearing the patch every day.
Besides, 
there might also be some errors in labelling as well.

A set of possible noisy labels are presented below: 

For example, 
Fig. \ref{fig_3} shows the standard waveform of type V premature ventricular contraction~\cite{r3}.

\begin{figure}[h]
\centering
\includegraphics[width=8cm]{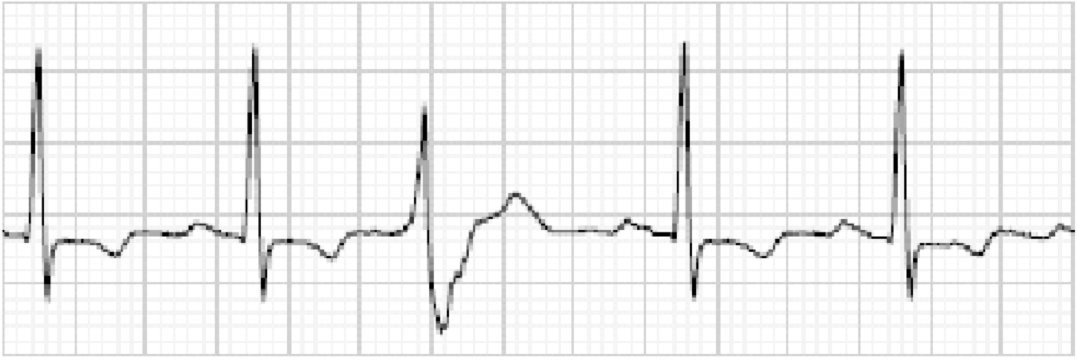}
\caption{Standard waveform of type V}
\label{fig_3}
\end{figure}

\begin{figure}[h]
     \centering
     \includegraphics[width=6cm]{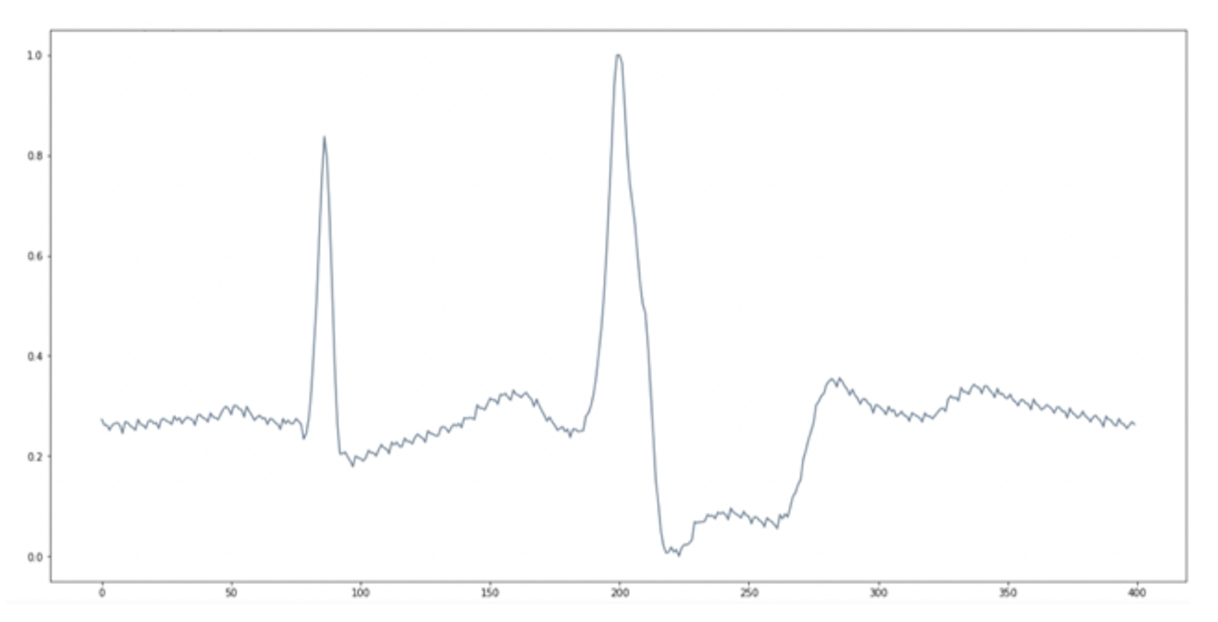}
     \caption{Example of "Clean label" in the dataset}
      \label{fig_4A}
\end{figure}

In our dataset, 
there are a few signals that have very similar features as shown in Fig. \ref{fig_4A}.
As such, 
we would consider it as a "clean label".

\begin{figure}[h]
     \centering
     \includegraphics[width=6cm]{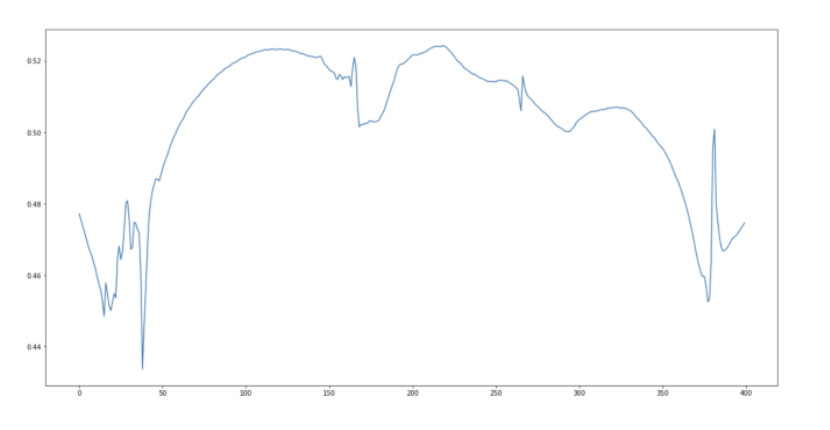}
     \caption{Example of "Noisy label" in the dataset}
      \label{fig_4B}
\end{figure}

However, 
there are also a few "noisy labels" in this dataset, 
such as the one shown in Fig.~\ref{fig_4B}. 
It was labelled as type V, 
but it's revised to be other type because it doesn't show the significant features of type V (Ventricular premature beat). 

\subsection{Data Segmentation}
To identify the QRS position, 
both low pass filtering and high pass filtering are used to mitigate the noise and locate the QRS. 
Low pass filters can be used to cancel the high pass noise efficiently while the high pass filters can limit the interference of P and T wave. After filtering using low pass and high pass, we used hand-crafted features to identify the QRS wave respectively.

\subsection{Data Pre-processing}
Data normalization is an essential step before feeding data into neural network for training. 
In deep learning, 
data are usually normalized to a mean of close to zero, 
which will usually speed up the learning process and result in a faster convergence. 
The reasons are as follows.

We define the binary cross-entropy loss as,
\begin{equation}
L=Y\times log(W\cdot X )+(1-Y)\log(1-W\cdot X)
\end{equation}
where $Y$ is the training label vector, $X$ is the input vector and $W$ is the weight matrix. 
If we take the partial derivative with respect to each weight, we have,  
\begin{equation}
\frac{\partial L}{\partial w_i}=x_i\times (a-y)
\label{eq2}
\end{equation}
where $x_i$ is corresponding activation output of this weight, $a$ is the network output probability and $y$ is the training label.

Weights update according to the following, 
\begin{equation}
w_i:= w_i - \alpha \times \frac{\partial L}{\partial w_i}
\label{eq3}
\end{equation}
where $\alpha$ is the learning rate.

Even though magnitudes are different, 
weights will be updated with some dependence on each other. 
And that usually indicates a slow convergence, 
because independent weight updates ensure all possible combinations of weights, 
which is crucial for approaching a local minimum. 

Therefore, 
in this work, 
we normalized our data to the range of $-0.5$ to $0.5$ according to Eq. (\ref{eq4}), 
which also makes sure the scale of our input data is consistent, 
suppressing the outliers. 
\begin{equation}
    x_i=\frac{x_i-\frac{\left\{\max{\left(x_i\right)}+\min{\left(x_i\right)}\right\}}{2}}{\max{\left(x_i\right)}-\min{\left(x_i\right)}}
    \label{eq4}
\end{equation}

\section{METHODOLOGY}\label{s3}
In this section, 
we discuss the methodology of our work. 

To implement this training scheme, there are four steps we need to do:\\
1. Find noisy signals by performing Confidence-level-based Learning;\\
2. Generate a complementary labels for noisy labels;\\
3. Perform negative learning by flipping the loss and gradient;\\
4. Change the outcome from "NOT Type X" into "Type X".

\subsection{Negative Learning and Confidence-level-based Learning}

\subsubsection{Positive Training}
Positive training is the traditional way of training a neural network, 
where the neural network is trained to learn the distribution of a given dataset and corresponding labels~\cite{r28}. 
For example, 
for a neural network to learn what is an image of a cat, 
we would “show” many pictures of cat and “tell” the neural network that “it’s a cat” multiple times until the neural network can recognize cats on its own. 

\subsubsection{Negative Training}
Negative training is the opposite of positive training. 
Instead of telling the neural network “it is a cat”, 
we tell it “it is not a dog”. 
This idea was first proposed in~\cite{r29} and used for increasing the accuracy of the label and decreasing the noisy labels. 
The labels used for negative training processes are called negative labels in this work.

\subsubsection{Example}
Usually, 
negative labels are more likely to be correctly labelled compared to positive labels under the same circumstances. 
This is because the accuracy of deciding whether an object belongs to a label is lower than deciding whether it doesn't belong to the label.
For example, 
there are totally 6 types of animals in a given dataset.
A cat image is mislabelled as a dog image accidentally, 
which might impair the training process. 
However, 
negative labelling process might give it a "not a monkey" label, 
which is still more likely to be correct($P_{correct}$= 5/6).

\subsubsection{Generate Complementary Labels }
As such, 
we can generate correct negative labels from positive labels even if it is not labelled correctly.

\begin{figure}[h]
\centering
\includegraphics[width=8cm]{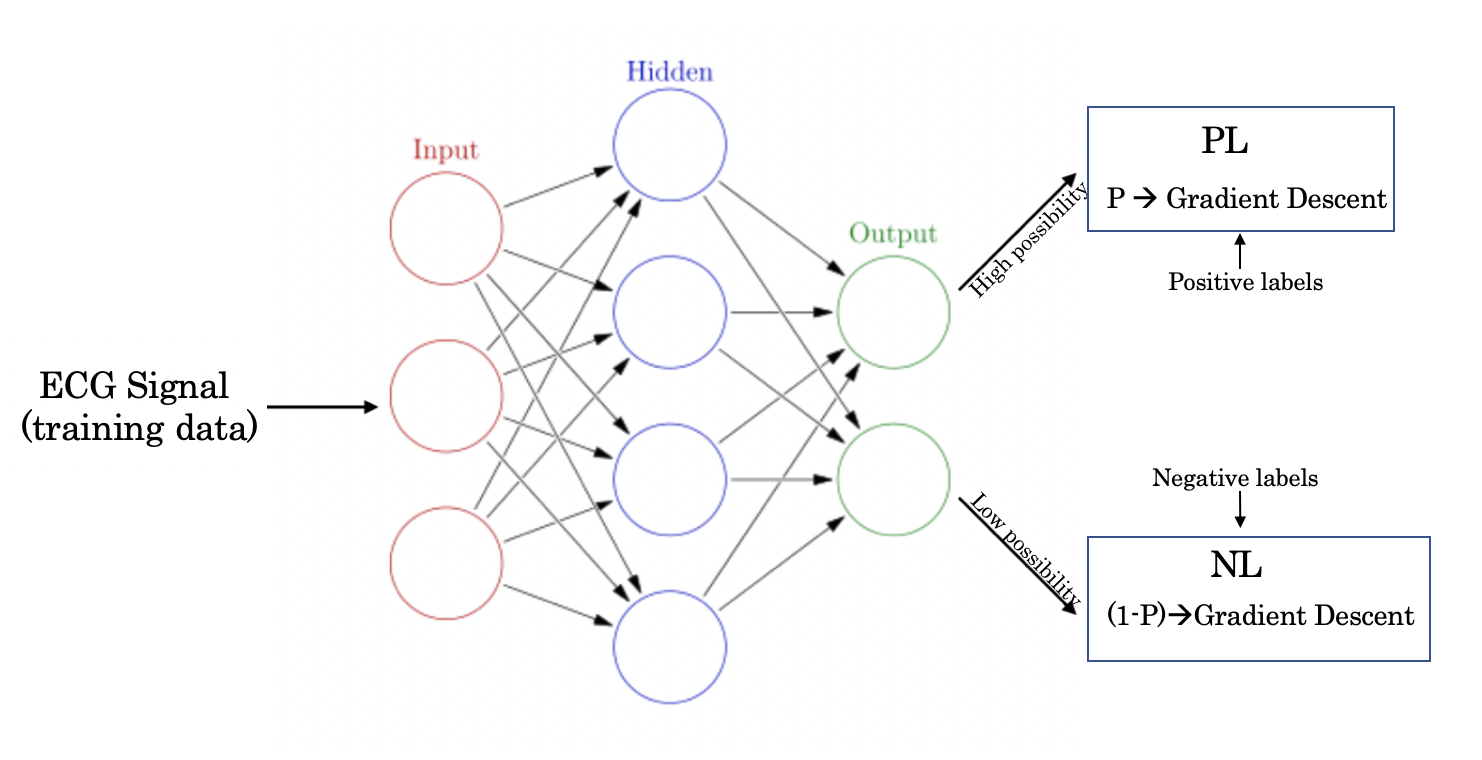}
\caption{Training process of Positive Learning and Negative Learning}`	
\end{figure}

Therefore, 
we can use a combination of positive training and negative training to improve the network performance when the training dataset is noisy. 
The authors of~\cite{r28} showed the performance of this training system on an ImageNet dataset with randomly added Gaussian noise.

To implement negative learning, 
\textbf{a complementary label $y'$ needs to be generated in the first place,
which means the input doesn't belong to the complementary label $y'$}. 
For example, if the data was labelled as Type 1 and its confidence level is relatively low. 
Then we will find its complementary label by randomly select a type from Type 2 - Type 6.
For instance, we select a Type 4 in this case. 
And the probability of "this signal is NOT Type 4" is 5/6, 
hence the probability of "this signal is Type 4" is 1/6.

\begin{figure}[h]
\centering
\includegraphics[width=8cm]{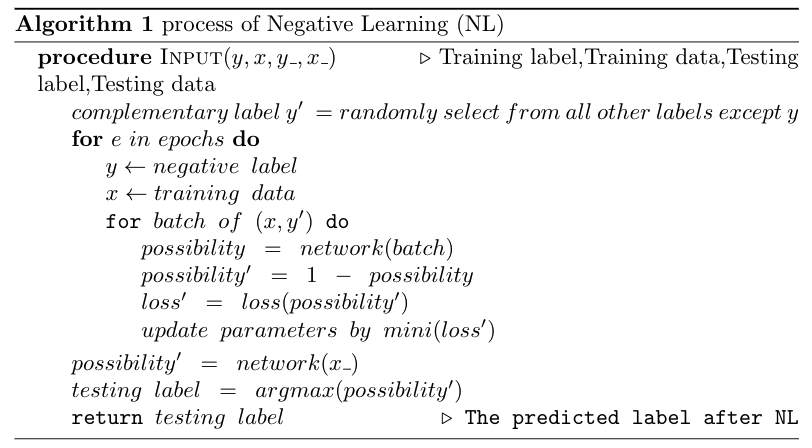}
\caption{Algorithm of Negative Training}
\label{fig_negtr}
\end{figure}

The detailed algorithm of generating a complementary label and negative label is shown in Fig. \ref{fig_negtr}.

\subsubsection{Negative Training implementation}

In the implementation, 
\textbf{a negative version of the cross-entropy loss function is used, }
in which the output possibility is flipped. 
For example, 
for a specific type, 
if the output possibility is $0.8$ in a regular cross-entropy loss measurement, 
the flipped one will be $0.2$. 
In implementation, 
the "not monkey" training is realized by reverting the activation function at the output layer, 
such that when the gradient is passed through layers, 
the model will fit the opposite way as opposed to positive learning. 

In more detail, 
if we want to train a "not monkey" label, 
we first calculate the output activation function for a label of "monkey",
which contains the information on how the weights affect the network output for a label of "monkey". 
Then, 
we take the negative of the output activation function, 
which holds the information on how the weights affect the network output for a label of "not monkey".
Hence,
passing gradients calculated in this way will result in fitting the opposite way of "monkey", 
i.e. "not monkey", 
yield a negative learning method.
By doing this, 
we can 'trick' the training process to negatively fit our negative labels. 
Hence, 
we can get our negative label $y\_$ here. 
In the implementation, 
we don't need to expand the labelling by considering other negative cases.

\textbf{The main difference between positive training and negative training is the loss function. }

Generally, 
when training with positive learning, 
the expression of cross entropy loss function is given by:
\begin{equation}
L(f,y) = - \sum_{k=1}^{n}{y_k\times \log\left(p_k\right)}
\end{equation}
where $L$ denotes the cross entropy loss, 
$n$ is the number of class, 
$y_k$ denotes current digit of the training label (which assumes to be one-hot encoding), 
and $p_k$ denotes output probability of current class.
The equation measures the performance of a classification model whose output is a probability value between $0$ and $1$. 
It calculates and accumulates the difference between predicted and actual labels.

Here, 
we sum up the loss of each type by calculating the product of $y_k$ and $\log {(p_k)}$.

However, 
negative training has exactly opposite loss function, where the output probability is reverted.

The expression of the cross entropy loss function is 
\begin{equation}
L(f,y\_) = - \sum_{k=1}^{n}{(y\_)_k\times \log\left(1-p_k\right)}
\end{equation}

\subsection{ResNet and Residual block}
Residual block is the building block of our proposed model. 
It contains a normal convolutional layer, 
except for a residual connection from the input of the block to output of the block. 
It was first proposed in~\cite{r30} as a solution for vanishing gradient when neural network goes deeper. 
Similar to~\cite{r30}, 
let’s consider $M(x)$ as the underlying mapping which the residual block is supposed to learn and $x$ denotes input to the residual block. 
The idea of residual learning is that rather than learning the underlying mapping $M(x)$, 
we learn the residual function $F(x)=M(x)-x$. 
And the underlying mapping can be reconstructed by an identity connection $F(x)+x$, 
which simply adds input to the output.

\begin{figure}[h]
\centering
\includegraphics[width=5cm]{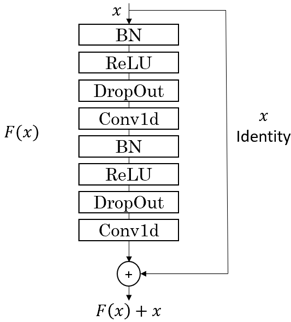}
\caption{Residual Block}
\label{fig_7}
\end{figure}

It has been proved that residual function F(x) is easier to learn and with extra information passed from input~\cite{r30}, 
the model can be trained deeper and deeper. 
If the output dimension is different from the input dimension, 
for example filter number changes, 
a linear projection $W_s$ needs to perform at the identity connection as $F\left(x\right)+W_s\times x$ (where  $W_s$ is a matrix with a dimension of a$\times$b), ensuring the identity connection’s dimension matches the output dimension, such that we can perform an element-wise addition.

Fig. \ref{fig_7} shows the structure of our proposed residual block. 
It consists of two 1-D convolution layers followed by Batch Normalization, 
ReLu activation and Dropout which improve the convergence of the training process. 
As can be seen, 
an identity connection is established from the input to the output with a simple addition operation.

\section{Experimental Setup}\label{s4}

\subsection{Neural Network Backbone Model}
\begin{figure}[h]
\centering
\includegraphics[width=9cm]{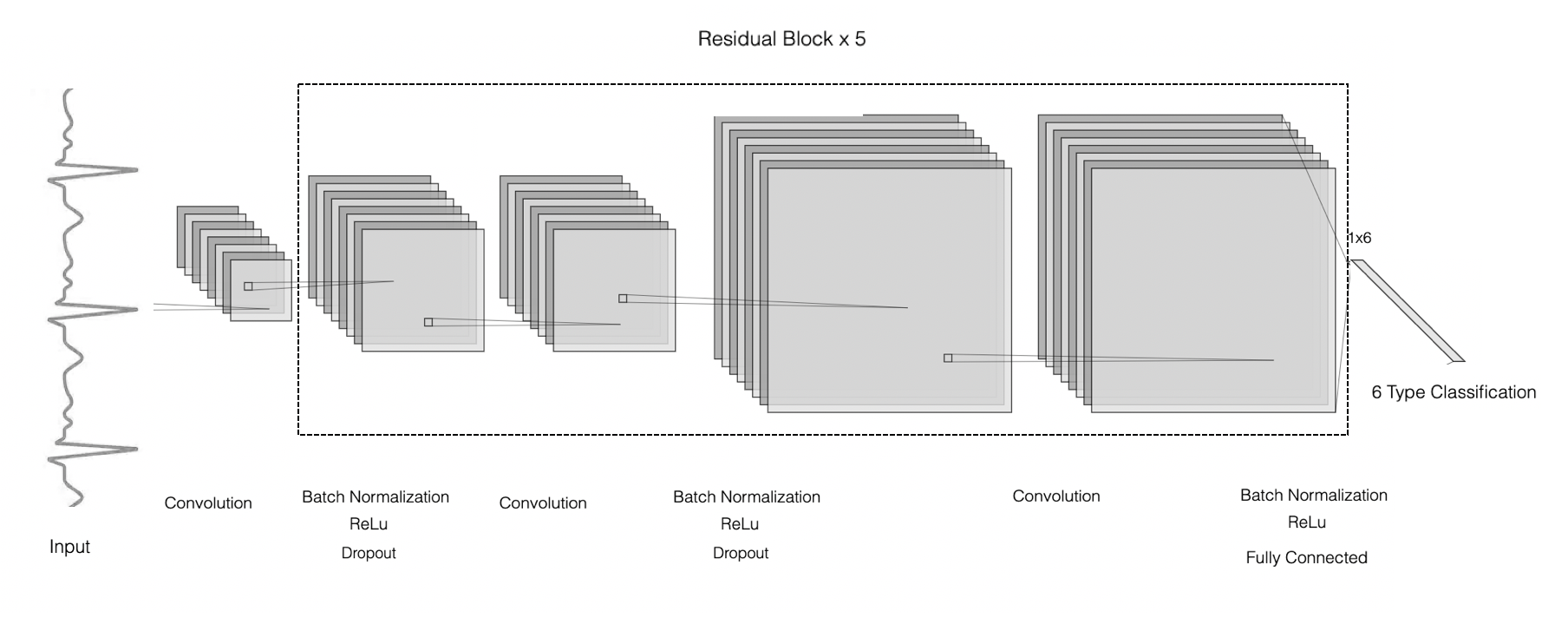}
\caption{Network Structure of Negative-ResNet}
\label{fig8}
\end{figure}

This backbone is a residual block based deep 1-D Convolution neural network which consists of 12 convolutional layers as shown in Fig. \ref{fig8}. 
Residual blocks which consist of a novel residual connection from the input to the output propagating information to deeper layers are basic building blocks of this backbone model~\cite{r30}. 
The number of filters in the residual block is double every four residual block and a max-pooling operation is performed on the residual connection such that the input is sub-sampled by half before adding to its output for every two residual blocks.

\begin{figure}[h]
\centering
\includegraphics[width=6cm]{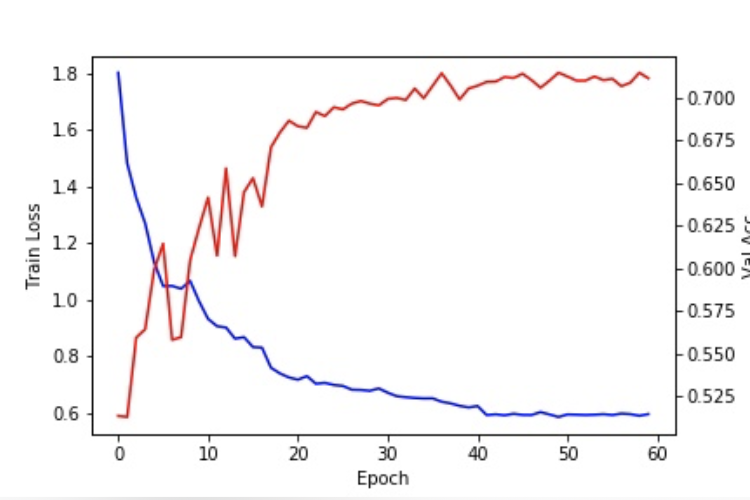}
\caption{Baseline Model Training Accuracy and Loss}
\label{fig9}
\end{figure}

We have constructed a baseline model with the same structure but normal positive training scheme. 
The result of baseline model is shown in \ref{fig9}. The highest accuracy achieved at convergence is $71\%$.

\subsection{Training Process}

\begin{figure}[h]
\centering
\includegraphics[width=8cm]{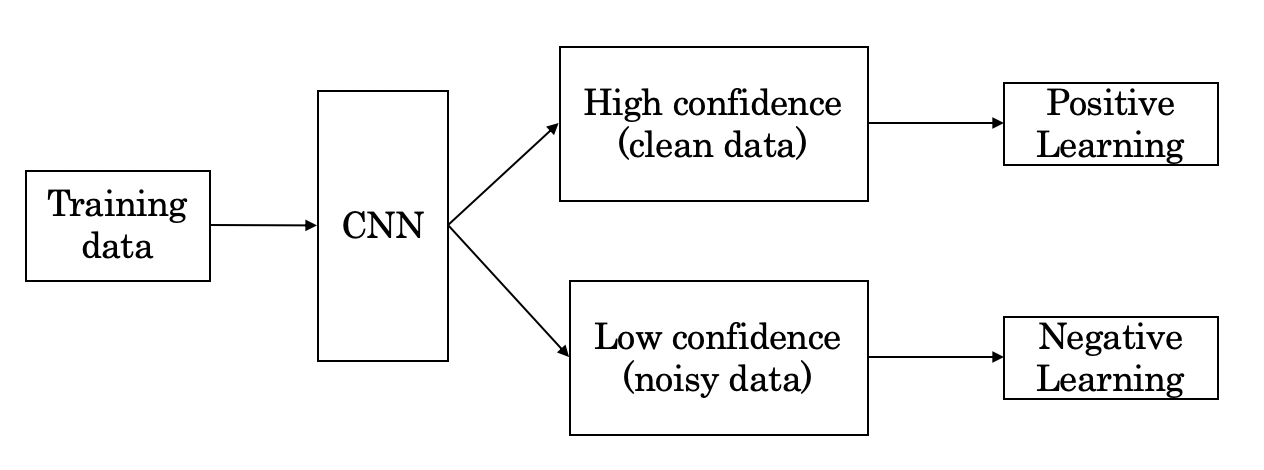}
\caption{Training process of Negative-ResNet}
\label{fig10}
\end{figure}

The whole training process of the system is shown in Fig. \ref{fig10}. 
The raw data is firstly feed into the CNN. 
Later on, 
the dataset is divided into clean data and noisy data based on the confidence level. 
Here, 
the confidence level refers to the output probability of the model. 
If the output confidence level is higher than a certain threshold, 
it is considered as clean data, vice versa. 

The clean data (high confidence level) would be trained positively, 
and the noisy data (low confidence level) would go through negative learning process. 
Loss calculated from both positive and negative learning will be added together, 
and then passed back for gradient descent. 
Weights will be updated at the same time. 

Different confidence levels are tested in our experiments. 
As can be seen from the results of Table \ref{tabIII}, 
the best result can be achieved when the confidence level is set to $80\%$, 
which is around $10\%$ higher than the baseline model. 
The confidence level controls how many data samples should be trained positively and how many data should be trained negatively. 
Therefore, 
it reflects the level of noisy labels contained in the dataset. 
And the best confidence level varies depending on different datasets. 

Note that originally there are six labels which consist of two interference labels: Type Q and Type E.

\begin{table}[h]
	\caption{Accuracy of Negative-ResNet with different confidence level}
	\begin{center}
		\begin{tabular}{|c|c|}
			\hline
			Confidence Level&Accuracy \\
			\hline
			99\%&~83.0\%\\
			90\%&~87.3\%\\
			80\%&~90.7\%\\
			70\%&~88.4\%\\
			60\%&~87.2\%\\
			50\%&~85.0\%\\
			40\%&~83.1\%\\
			30\%&~83.1\%\\
			\hline
		\end{tabular}
	\end{center}
	\label{tabIII}
\end{table}

As such, 
the accuracy reaches $81\%$ when training with $80\%$ confidence level. 
The accuracy and loss plot of training with $80\%$ of confidence level is shown \label{fig11}.

Note: The confidence level for Type A is set to be 50\% such that few Type A sample will go through the Negative Learning. It is because Type A in our dataset is confirmed by our medical expert to be of high labeling correctness. 

\begin{figure}[h]
\centering
\includegraphics[width=6cm]{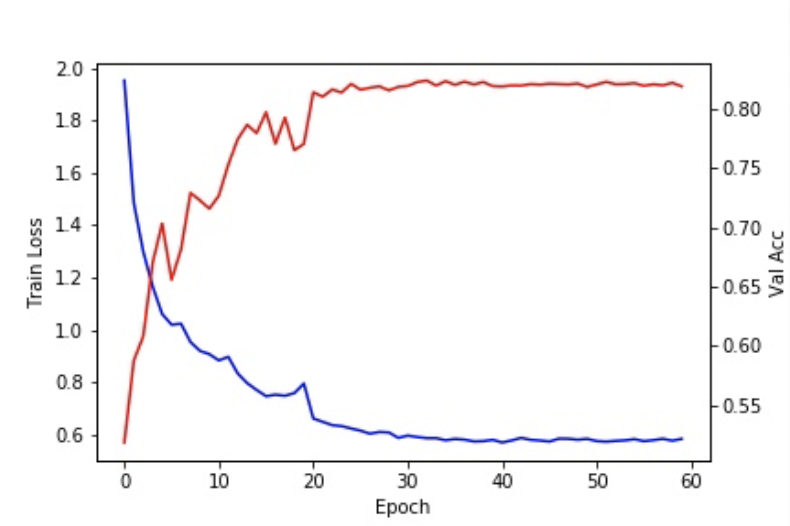}
\caption{Training Accuracy and Loss Plot of 80\% Confidence Level}
\label{fig11}
\end{figure}

\section{ASSESSMENT CRITERION AND RESULT ANALYSIS}
\label{s5}
In this work, 
we use accuracy, 
precision, 
and recall to evaluate the performance of our classifier. 
These are the commonly used and standardized parameters for machine learning and deep learning. 
The performance of the model can be determined by these parameters since it can not only show how precise the model is, 
but also whether the model is biased.

In the confusion matrix below, 
Type Q and Type E are combined as Type Q.

The definitions of these parameters are listed below, 
where TP stands for True Positive, 
FP stands for False Positive, 
and FN stands for False Negative:
\begin{equation}
precision=\frac{TP}{TP+FP};
\end{equation}

\begin{equation}
recall=\frac{TP}{TP+FN};
\end{equation}

\begin{equation}
F1 score=2 * \frac{precision \times recall}{precision + recall}
\end{equation}
Here, 
we also used $K$-Fold cross-validation to validate our results. 
$N$ is chosen to be two. 
In more detail, 
two patients will be selected randomly and the remaining patients will be used as training data. 

\subsection{Baseline Model}
\begin{figure}[h]
\centering
\includegraphics[width=7cm]{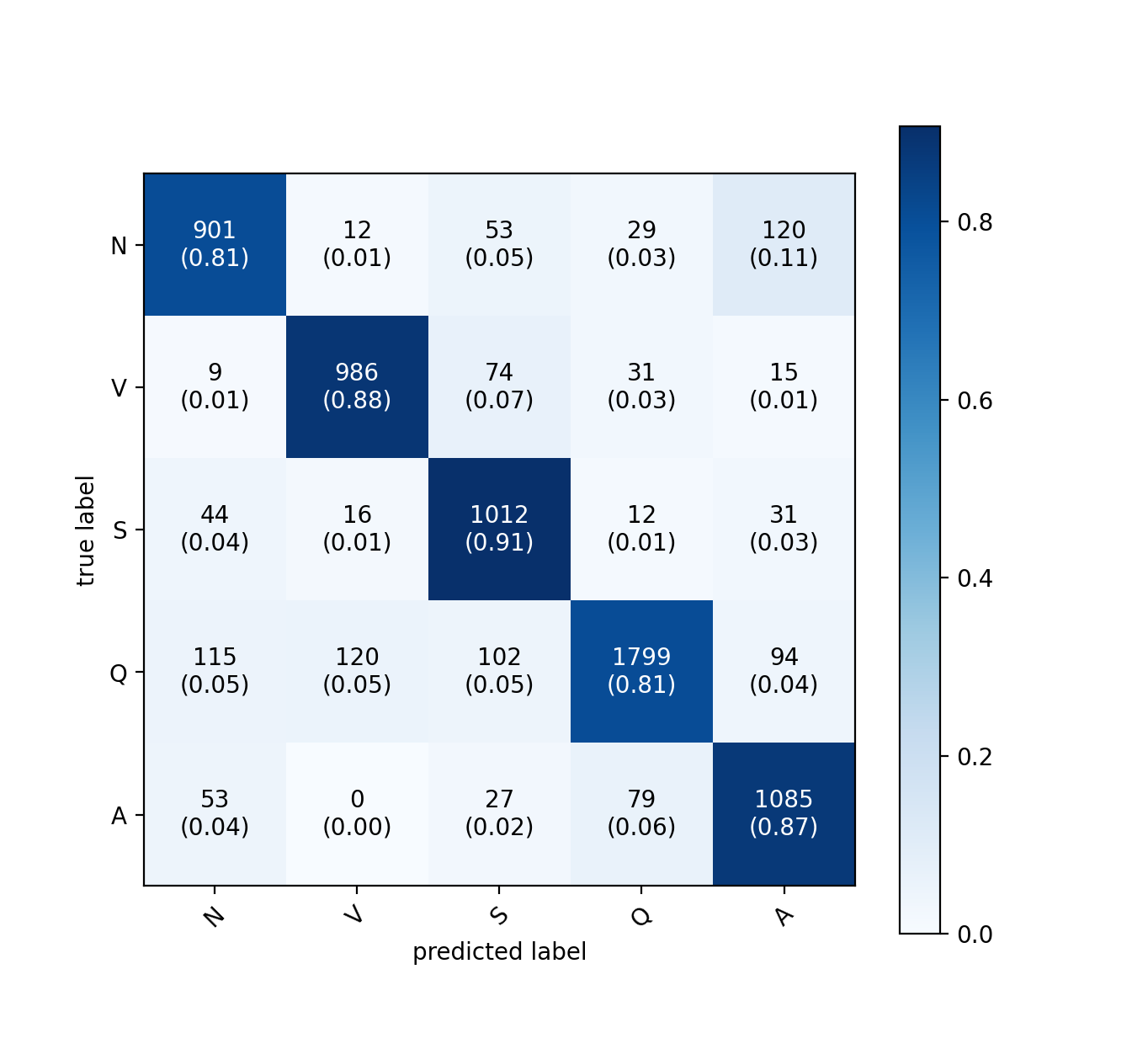}
\caption{Confusion Matrix of Baseline Model}
\label{fig12}
\end{figure}

The confusion matrix for the baseline model is shown in Fig. \ref{fig12}. 
The precision and recall of the baseline model are shown in Table \ref{tab4}.

\begin{table}[h]
\caption{Precision and Recall for Baseline Model}
	\begin{center}
		\begin{tabular}{|c|c|c|c|c|c|}
			\hline
			  &N&V&S&Q&A\\
			\hline
		    mean Precision&0.80&0.87&0.79&0.92& 0.81\\
			mean Recall&0.81& 0.88& 0.91&0.81& 0.87\\
            mean F1 score&0.80&0.88&0.85&0.86&0.84\\    
			\hline
		\end{tabular}
	\end{center}
	\label{tab4}
\end{table}

\subsection{Negative ResNet}
\begin{figure}[h]
\centering
\includegraphics[width=8cm]{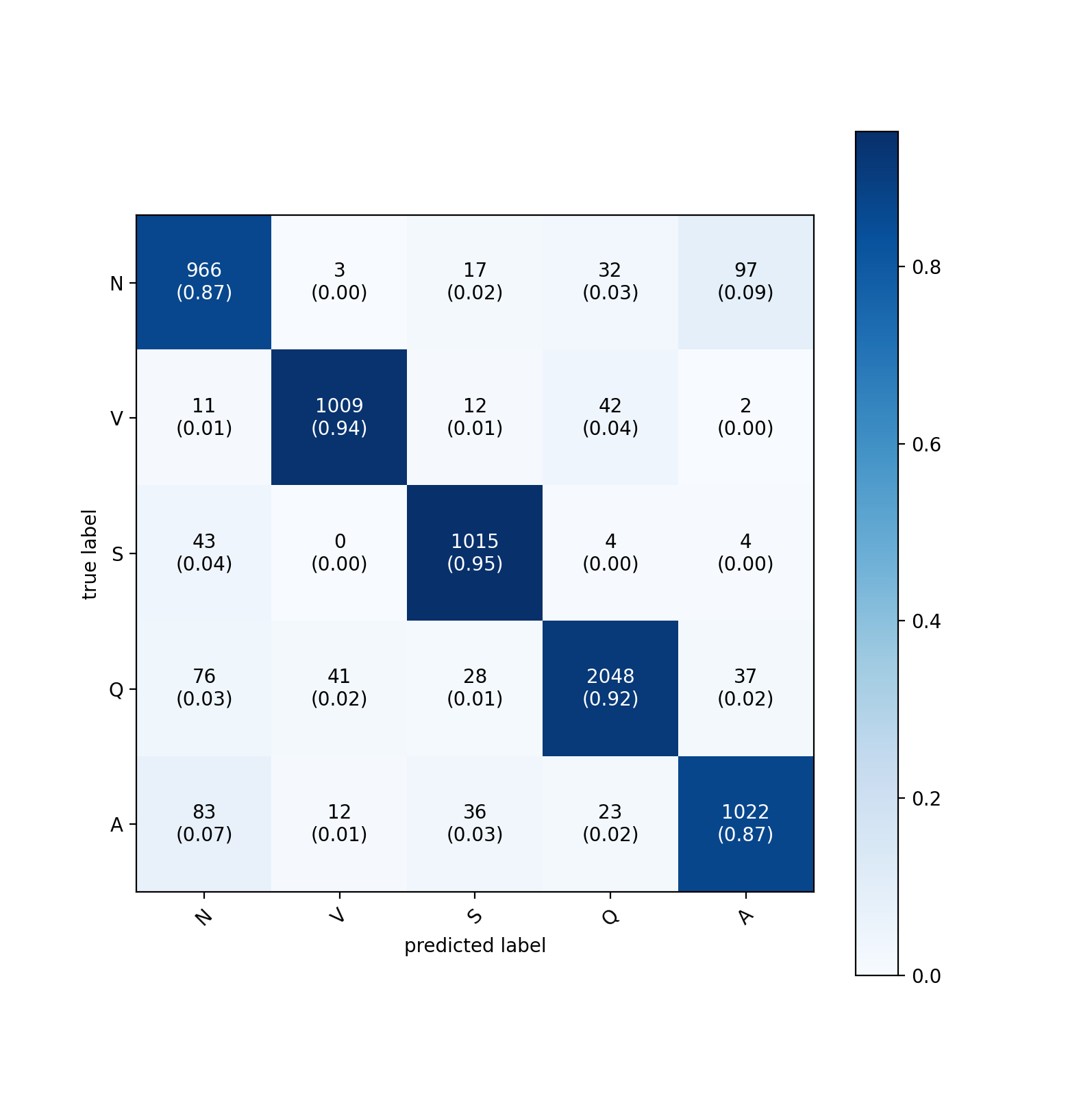}
\caption{Confusion Matrix of Negative ResNet}
\label{fig13}

\end{figure}

The confusion matrix, 
precision and recall of optimal Negative ResNet are shown in Fig. \ref{fig13} and Table \ref{tab5}, respectively.

\begin{table}[h]
\caption{Precision and Recall for Negative ResNet}
	\begin{center}
		\begin{tabular}{|c|c|c|c|c|c|}
			\hline
			  &N&V&S&Q&A\\
			\hline
		    mean Precision&0.82 & 0.95& 0.92& 0.95& 0.88\\
			mean Recall&0.86& 0.94 &0.95& 0.92& 0.87\\
            mean F1 score&0.84&0.94&0.93&0.94&0.87\\      
			\hline
		\end{tabular}
	\end{center}
	\label{tab5}
\end{table}

Generally speaking, 
the result of negative training is $4-11\%$ better than the normal training process, 
especially for interference, 
where an increment of $11\%$ accuracy can be achieved. 

However, for type A, 
the accuracy remains the same when going through negative training. 
This is mainly because for type A (Atrial fibrillation), 
the original labelling process is relatively accurate. 
As such, using negative learning in this case will not have a significant improvement in the accuracy of type A. 
It proves our idea that negative learning performs better in a noisy dataset.

\section{CONCLUSION AND FUTURE DIRECTION }
\label{s6}

This work introduced a single-lead ambulatory ECG patch iRealCare, 
which was used to construct a large real-patient dataset. 
In addition, a residual block based neural network with a novel negative training scheme was proposed to train an ECG classifier from this dataset. 
Even with a considerable amount of erroneous labels in the dataset, 
our proposed training scheme showed a satisfactory and better result compared to the baseline model, a 6.2\% of increase in accuracy was observed and the overall accuracy reached 91.0\%. 

Our proposed training scheme is useful in reducing the reliance on quality of the training dataset, such that the training performance can be improved and it can be extended to more applications. 

Further down this research, we believe that neural network and our proposed training scheme can not only applied into  the cardiology diagnosis process, it can improved other medical diagnosis processes in a similar manner as well. We will keep on exploring towards this direction and hopefully applying it on other biomedical signal processing field.


\begin{thebibliography}{00}

\bibitem{r8}S. Yu and Y. Chen, "Electrocardiogram beat classification based on wavelet transformation and probabilistic neural network", Pattern Recognition Letters, vol. 28, no. 10, pp. 1142-1150, 2007. Available: 10.1016/j.patrec.2007.01.017.


\bibitem{r25}Y. LeCun, Y. Bengio and G. Hinton, "Deep learning", Nature, vol. 521, no. 7553, pp. 436-444, 2015. Available: 10.1038/nature14539

\bibitem{r26} B. C. Russell, A. Torralba, K. P. Murphy, and W. T. Freeman, "LabelMe: a database and web-based tool for image annotation," International journal of computer vision, vol. 77, no. 1-3, pp. 157-173, 2008.

\bibitem{r1} H. Huikuri, A. Castellanos and R. Myerburg, "Sudden Death Due to Cardiac Arrhythmias", New England Journal of Medicine, vol. 345, no. 20, pp. 1473-1482, 2001. 

\bibitem{r2}N. El-Sherif et al., "Definition of the best prediction criteria of the time domain signal-averaged electrocardiogram for serious arryhthmic events in the postinfarction period," Journal of the American College of Cardiology, vol. 25, no. 4, pp. 908-914, 1995.

\bibitem{r3}C. Ramanathan, R. N. Ghanem, P. Jia, K. Ryu, and Y. Rudy, "Noninvasive electrocardiographic imaging for cardiac electrophysiology and arrhythmia," Nature medicine, vol. 10, no. 4, pp. 422-428, 2004.

\bibitem{r4}T. G. Farrell et al., "Risk stratification for arrhythmic events in postinfarction patients based on heart rate variability, ambulatory electrocardiographic variables and the signal-averaged electrocardiogram," Journal of the American College of Cardiology, vol. 18, no. 3, pp. 687-697, 1991.

\bibitem{r5}Y. Yeh and W. Wang, "QRS complexes detection for ECG signal: The Difference Operation Method", Computer Methods and Programs in Biomedicine, vol. 91, no. 3, pp. 245-254, 2008.

\bibitem{r6}A. L. Goldberger, Z. D. Goldberger, and A. Shvilkin, Clinical electrocardiography: a simplified approach e-book. Elsevier Health Sciences, 2017.


\bibitem{r7} P. Addison, J. Watson, G. Clegg, M. Holzer, F. Sterz and C. Robertson, "Evaluating arrhythmias in ECG signals using wavelet transforms", IEEE Engineering in Medicine and Biology Magazine, vol. 19, no. 5, pp. 104-109, 2000. Available: 10.1109/51.870237.




\bibitem{r9}M. Elgendi, A. Mohamed, and R. Ward, "Efficient ECG compression and QRS detection for e-health applications," Scientific reports, vol. 7, no. 1, pp. 1-16, 2017.

\bibitem{r10} N. Thakor and Y. Zhu, "Applications of adaptive filtering to ECG analysis: noise cancellation and arrhythmia detection", IEEE Transactions on Biomedical Engineering, vol. 38, no. 8, pp. 785-794, 1991. Available: 10.1109/10.83591.

\bibitem{r11}C. Venkatesan, P. Karthigaikumar, A. Paul, S. Satheeskumaran, and R. Kumar, "ECG signal preprocessing and SVM classifier-based abnormality detection in remote healthcare applications," IEEE Access, vol. 6, pp. 9767-9773, 2018.

\bibitem{r12}J. Deng, W. Dong, R. Socher, L. Li, Kai Li and Li Fei-Fei, "ImageNet: A large-scale hierarchical image database," 2009 IEEE Conference on Computer Vision and Pattern Recognition, Miami, FL, 2009, pp. 248-255.

\bibitem{r13}A. Majumdar and R. Ward, "Robust greedy deep dictionary learning for ECG arrhythmia classification," 2017 International Joint Conference on Neural Networks (IJCNN), Anchorage, AK, 2017, pp. 4400-4407.

\bibitem{r14} S. Kiranyaz, T. Ince and M. Gabbouj, "Real-Time Patient-Specific ECG Classification by 1-D Convolutional Neural Networks", IEEE Transactions on Biomedical Engineering, vol. 63, no. 3, pp. 664-675, 2016. Available: 10.1109/tbme.2015.2468589.

\bibitem{r15}  Acharya, U. Rajendra, et al. "A deep convolutional neural network model to classify heartbeats." Computers in biology and medicine 89 (2017): 389-396.

\bibitem{r16} S. Singh, S. K. Pandey, U. Pawar, and R. R. Janghel, "Classification of ECG arrhythmia using recurrent neural networks," Procedia computer science, vol. 132, pp. 1290-1297, 2018.

\bibitem{r17} Ö. Yildirim, "A novel wavelet sequence based on deep bidirectional LSTM network model for ECG signal classification," Computers in biology and medicine, vol. 96, pp. 189-202, 2018.

\bibitem{r18}S. L. Oh, E. Y. Ng, R. San Tan, and U. R. Acharya, "Automated diagnosis of arrhythmia using combination of CNN and LSTM techniques with variable length heart beats," Computers in biology and medicine, vol. 102, pp. 278-287, 2018.

\bibitem{r19} O. Yildirim, U. B. Baloglu, R.-S. Tan, E. J. Ciaccio, and U. R. Acharya, "A new approach for arrhythmia classification using deep coded features and LSTM networks," Computer methods and programs in biomedicine, vol. 176, pp. 121-133, 2019.

\bibitem{r20}M. Salem, S. Taheri and J. Yuan, "ECG Arrhythmia Classification Using Transfer Learning from 2- Dimensional Deep CNN Features," 2018 IEEE Biomedical Circuits and Systems Conference (BioCAS), Cleveland, OH, 2018, pp. 1-4.

\bibitem{r21}S. Kiranyaz, T. Ince and M. Gabbouj, "Real-Time Patient-Specific ECG Classification by 1-D Convolutional Neural Networks", IEEE Transactions on Biomedical Engineering, vol. 63, no. 3, pp. 664-675, 2016. Available: 10.1109/tbme.2015.2468589.

\bibitem{r22} G. Moody and R. Mark, "The impact of the MIT-BIH Arrhythmia Database", IEEE Engineering in Medicine and Biology Magazine, vol. 20, no. 3, pp. 45-50, 2001. Available: 10.1109/51.932724.

\bibitem{r23} T. A. Nappholz, W. N. Hursta, A. K. Dawson, and B. M. Steinhaus, "Implantable ambulatory electrocardiogram monitor," ed: Google Patents, 1992.

\bibitem{r24}V. K. Yeragani et al., "Decreased heart-period variability in patients with panic disorder: a study of Holter ECG records," Psychiatry research, vol. 78, no. 1-2, pp. 89-99, 1998.


\bibitem{01}. Algan and I. Ulusoy, “Image classification with deep learning in the
presence of noisy labels: A survey,” arXiv preprint arXiv:1912.05170,
2019.

\bibitem{02}B. Frenay and M. Verleysen, “Classification in the presence of label ´
noise: a survey,” IEEE transactions on neural networks and learning
systems, vol. 25, no. 5, pp. 845–869, 2014.

\bibitem{03} Patrini, A. Rozza, A. Krishna Menon, R. Nock, and L. Qu, “Making
deep neural networks robust to label noise: A loss correction approach,”
in Proceedings of the IEEE Conference on Computer Vision and Pattern
Recognition, 2017, pp. 1944–1952.

\bibitem{04} D. Hendrycks, M. Mazeika, D. Wilson, and K. Gimpel, “Using trusted
data to train deep networks on labels corrupted by severe noise,” in
Advances in neural information processing systems, 2018, pp. 10 456–
10 465.

\bibitem{05} X. Chen and A. Gupta, “Webly supervised learning of convolutional
networks,” in Proceedings of the IEEE International Conference on
Computer Vision, 2015, pp. 1431–1439.

\bibitem{06} A. J. Bekker and J. Goldberger, “Training deep neural-networks based
on unreliable labels,” in 2016 IEEE International Conference on
Acoustics, Speech and Signal Processing (ICASSP). IEEE, 2016, pp.
2682–2686.

\bibitem{07} J. Goldberger and E. Ben-Reuven, “Training deep neural-networks
using a noise adaptation layer,” 2016.

\bibitem{08} S. Sukhbaatar and R. Fergus, “Learning from noisy labels with deep
neural networks,” arXiv preprint arXiv:1406.2080, vol. 2, no. 3, p. 4,
2014.

\bibitem{09} B. Frenay and M. Verleysen, “Classification in the presence of label ´
noise: a survey,” IEEE transactions on neural networks and learning
systems, vol. 25, no. 5, pp. 845–869, 2014.

\bibitem{10} J. Krause, B. Sapp, A. Howard, H. Zhou, A. Toshev, T. Duerig,
J. Philbin, and L. Fei-Fei, “The unreasonable effectiveness of noisy data
for fine-grained recognition,” in European Conference on Computer
Vision. Springer, 2016, pp. 301–320.

\bibitem{11} C. G. Northcutt, T. Wu, and I. L. Chuang, “Learning with confident
examples: Rank pruning for robust classification with noisy labels,”
in Uncertainty in Artificial Intelligence - Proceedings of the 33rd
Conference, UAI 2017, may 2017.

\bibitem{12} X. Wu, R. He, Z. Sun, and T. Tan, “A light CNN for deep face
representation with noisy labels,” IEEE Transactions on Information
Forensics and Security, vol. 13, no. 11, pp. 2884–2896, 2018.

\bibitem{13} J. Huang, L. Qu, R. Jia, and B. Zhao, “O2u-net: A simple noisy label
detection approach for deep neural networks,” in Proceedings of the
IEEE International Conference on Computer Vision, 2019, pp. 3326–
3334.

\bibitem{14} Y. Bengio, J. Louradour, R. Collobert, and J. Weston, “Curriculum
learning,” in Proceedings of the 26th annual international conference
on machine learning. ACM, 2009, pp. 41–48.

\bibitem{15} M. P. Kumar, B. Packer, and D. Koller, “Self-paced learning for
latent variable models,” in Advances in Neural Information Processing
Systems, 2010, pp. 1189–1197.

\bibitem{16} B. Han, I. W. Tsang, L. Chen, P. Y. Celina, and S.-F. Fung, “Progressive
stochastic learning for noisy labels,” IEEE transactions on neural
networks and learning systems, no. 99, pp. 1–13, 2018.

\bibitem{17} L. Jiang, Z. Zhou, T. Leung, L.-J. Li, and L. Fei-Fei, “Mentornet:
Learning data-driven curriculum for very deep neural networks on
corrupted labels,” arXiv preprint arXiv:1712.05055, 2017

\bibitem{18} M. Ren, W. Zeng, B. Yang, and R. Urtasun, “Learning to reweight
examples for robust deep learning,” arXiv preprint arXiv:1803.09050,
2018.

\bibitem{19} S. Jenni and P. Favaro, “Deep bilevel learning,” in Proceedings of the
European Conference on Computer Vision (ECCV), 2018, pp. 618–633.

\bibitem{20} J. Shu, Q. Xie, L. Yi, Q. Zhao, S. Zhou, Z. Xu, and D. Meng, “Metaweight-net: Learning an explicit mapping for sample weighting,” in
Advances in Neural Information Processing Systems, 2019, pp. 1917–
1928.

\bibitem{21} Y. Wang, W. Liu, X. Ma, J. Bailey, H. Zha, L. Song, and S.-T. Xia,
“Iterative learning with open-set noisy labels,” in Proceedings of the
IEEE Conference on Computer Vision and Pattern Recognition, 2018,
pp. 8688–8696.

\bibitem{22} N. Manwani and P. Sastry, “Noise tolerance under risk minimization,”
IEEE transactions on cybernetics, vol. 43, no. 3, pp. 1146–1151, 2013.

\bibitem{23} P. L. Bartlett, M. I. Jordan, and J. D. McAuliffe, “Convexity, classification, and risk bounds,” Journal of the American Statistical Association,
vol. 101, no. 473, pp. 138–156, 2006.

\bibitem{24} A. Ghosh, H. Kumar, and P. Sastry, “Robust loss functions under label
noise for deep neural networks,” in Thirty-First AAAI Conference on
Artificial Intelligence, 2017.

\bibitem{25} X. Wang, E. Kodirov, Y. Hua, and N. M. Robertson, “Improved
Mean Absolute Error for Learning Meaningful Patterns from Abnormal
Training Data,” Tech. Rep., 2019.

\bibitem{26} V. Mnih and G. E. Hinton, “Learning to label aerial images from noisy
data,” in Proceedings of the 29th International conference on machine
learning (ICML-12), 2012, pp. 567–574.

\bibitem{27} Y. Xu, P. Cao, Y. Kong, and Y. Wang, “L dmi: A novel informationtheoretic loss function for training deep nets robust to label noise,” in
Advances in Neural Information Processing Systems, 2019, pp. 6222–
6233.

\bibitem{28} B. Van Rooyen, A. Menon, and R. C. Williamson, “Learning with
symmetric label noise: The importance of being unhinged,” in Advances
in Neural Information Processing Systems, 2015, pp. 10–18.

\bibitem{29} L. P. Garcia, A. C. de Carvalho, and A. C. Lorena, “Noise detection in
the meta-learning level,” Neurocomputing, vol. 176, pp. 14–25, 2016.

\bibitem{30} B. Han, G. Niu, J. Yao, X. Yu, M. Xu, I. Tsang, and M. Sugiyama,
“Pumpout: A meta approach for robustly training deep neural networks
with noisy labels,” arXiv preprint arXiv:1809.11008, 2018.

\bibitem{31} J. Li, Y. Wong, Q. Zhao, and M. Kankanhalli, “Learning to learn from
noisy labeled data,” 2018.

\bibitem{32} N. Srivastava, G. Hinton, A. Krizhevsky, I. Sutskever, and R. Salakhutdinov, “Dropout: a simple way to prevent neural networks from
overfitting,” The journal of machine learning research, vol. 15, no. 1,
pp. 1929–1958, 2014.

\bibitem{33} I. J. Goodfellow, J. Shlens, and C. Szegedy, “Explaining and harnessing
adversarial examples,” arXiv preprint arXiv:1412.6572, 2014.

\bibitem{34} H. Zhang, M. Cisse, Y. N. Dauphin, and D. Lopez-Paz, “mixup: Beyond
Empirical Risk Minimization,” oct 2017.

\bibitem{35} X. Yu, T. Liu, M. Gong, and D. Tao, “Learning with biased complementary labels,” in Proceedings of the European Conference on Computer
Vision (ECCV), 2018, pp. 68–83.

\bibitem{36} W. Zhang, Y. Wang, and Y. Qiao, “Metacleaner: Learning to hallucinate clean representations for noisy-labeled visual recognition,” in
Proceedings of the IEEE Conference on Computer Vision and Pattern
Recognition, 2019, pp. 7373–7382



\bibitem{r27}J. Leng, "A Wearable ECG Monitor for Deep Learning Based Real-Time Cardiovascular Disease Detection", Available: https://www.sydneyiot.com/ecg-monitoring-module.

\bibitem{r28}Y. Kim, J. Yim, J. Yun and J. Kim, "NLNL: Negative Learning for Noisy Labels",  The IEEE International Conference on Computer Vision (ICCV), 2019, pp. 101-110

\bibitem{r29}S. Sukhbaatar, J. Bruna, M. Paluri, L. Bourdev and R. Fergus, "Training Convolutional Networks with Noisy Labels", arXiv:1406.2080v4 [cs.CV], 10 Apr 2015.

\bibitem{r30}K. He, X. Zhang, S. Ren and J. Sun, "Deep Residual Learning for Image Recognition",arXiv:1512.03385v1 [cs.CV], 10 Dec 2015.


\end{thebibliography}
\end{document}